\documentclass[journal=jacsat,manuscript=article]{achemso}
\usepackage{setspace}
\usepackage{graphicx}
\usepackage{amsmath} 
\usepackage{mathtools}
\usepackage{amsfonts}
\usepackage{bbm}
\usepackage{braket}
\usepackage{caption}
\usepackage{graphicx}
\usepackage{float}
\usepackage{caption}
\usepackage{subcaption}
\usepackage{url}
\usepackage{color}
\usepackage[ruled]{algorithm2e}

\usepackage{amssymb}
\usepackage{cancel}
\usepackage{upgreek}
\usepackage{tabularx}
\usepackage{color}
\usepackage{units}
\usepackage{hhline}
\usepackage{textfit} 

\usepackage{slashed}
\graphicspath{ {images/} }

\usepackage{graphicx}
\usepackage{dcolumn}
\usepackage{bm}
\usepackage{hyperref}
\usepackage[mathlines]{lineno}

\usepackage{tikz}
\usetikzlibrary{positioning}
\usepackage{qcircuit}
\usepackage{braket}
\usepackage{indentfirst}
\usepackage{float}

\usepackage[version=3]{mhchem} 

\author{Kumar J. B. Ghosh}
\email{jb.ghosh@outlook.com}
\affiliation{Department of Electrical and Computer Engineering,\\ University of Denver, Denver, CO, 80210, USA.
}%
\author{Sabre Kais}
\email{kais@purdue.edu}
\affiliation{ Department of Chemistry  and Physics,\\
Purdue University, West Lafayette, IN, 47906, USA.
}%
\author{Dudley R. Herschbach}
\email{dherschbach@gmail.com}
\affiliation{ Department of Chemistry and Chemical Biology,\\ Harvard University, Cambridge MA 02138, USA 
}%

\title[An \textsf{achemso} demo]
  {Dimensional Interpolation for Random Walk }

\abbreviations{IR,NMR,UV}

\begin{document}










\begin{abstract}
We employ a simple and accurate dimensional interpolation formula for the shapes of random walks at $D=3$ and $D=2$ based on the analytically known solutions at both limits $D=\infty$ and $D=1$. The results obtained for the radii of gyration of an arbitrary shaped object are about $2\%$ error compared with accurate numerical results at $D = 3$ and $D = 2$. We also calculated the asphericity for a three-dimensional random walk using the dimensional interpolation formula. Result agrees very well with the numerically simulated result. The method is general and can be used to estimate other properties of random walks. 

 
\end{abstract}

\maketitle




\section{Introduction}\label{sec: Introduction}

The random walk problem is introduced by Pearson in 1905 \cite{pearson1905problem} and the following developed theory successfully implemented in different branches of science and engineering, for e.g., hydrodynamics \cite{lebowitz1984nonequilibrium}, astronomy \cite{chandrasekhar1943stochastic}, chemistry \cite{domb1972cluster}, polymer sciences \cite{de1979scaling, vsolc1973statistical}, mathematics \cite{mandelbrot1982fractal}, biology \cite{codling2008random}, and economics \cite{cooper1982world}. Many other topics in condensed matter physics include percolation clusters \cite{essam1972phase}, lattice animals \cite{stauffer1979scaling, aronovitz1987universal}, disordered magnetic systems and spin glasses \cite{witten1986tenuous}, anisotropy, and random fractals \cite{family1985random}.  The studies exploring the shapes of random walks have appeared, e.g. \cite{kuhn1934gestalt, flory1953principles}.  Dimensional scaling offers simple solutions at $D = 1$ and $D \to \infty$ limits, then often interpolates to obtain accurate results for $D = 3$, in many areas of chemical physics \cite{herschbach2012dimensional, goodson1992large, zhen1993large, kais19941, rudnick1987shapes, loeser1991dimensional, kais1993dimensional, wei2008dimensional, wei2007dimensional, kais1994large, germann1993large}.   Recently, we have used the dimensional interpolation formula developed by one of us \cite{Herschbach2017} to obtain the ground state energies for few electron atoms, simple diatomic molecules \cite{ghosh2020unorthodox} and also extended systems like metallic hydrogen \cite{metallic_hydrogen}.  Rudnick et al studied the random walks in high spacial dimensions \cite{rudnick1987shapes, rudnick1987shapes2} and developed $1/D$ expansion to study the shape of a random walk in three dimensions. We employ a simple and accurate dimensional interpolation formula using dimensional limits $D = 1$ and $D \to \infty$ to analyze the random walk problem at $D=3$ and $D=2$, and obtain physical quantities like radius of gyration and asphericity describing an arbitrary shaped two- and three-dimensional objects. 


\section{\label{sec:Dimensional interpolation formula for random walk}   Basic dimensional interpolation formula} 

As a recapitulation, we start with the dimensional interpolation formula for atomic, molecular, and extended systems developed in our previous works. For dimensional scaling of atoms and molecules, the energy erupts to infinity as $D\to 1$ and vanishes as $D\to \infty$. Hence, we adopt scaled units (with hartree atomic units) whereby $E_D = \left( Z/\beta \right)^2 \epsilon_D$ and $\beta = \frac{1}{2} \left(D - 1 \right)$, so the reduced energy $\epsilon_D$ remains finite in both limits. 

The interpolation for atoms, developed in Ref. \cite{Herschbach2017}, weights the dimensional limits by $\delta = 1/D$, providing $\delta \epsilon_1$ and $\left( 1 - \delta \right) \epsilon_\infty$ in a simple analytic formula
\begin{equation}
\epsilon_D = \delta \epsilon_1 + \left( 1 - \delta \right) \epsilon_\infty + \left[ \epsilon_D ^{(1)} - \delta \epsilon_1^{(1)} - \left( 1 - \delta \right) \epsilon_\infty^{(1)} \right] \lambda . \label{interpolation_formula}
\end{equation}

For a diatomic molecule, a different scaling scheme is used and illustrated. The rescaling of distances is:

\begin{equation}
R \to \delta R^\prime \text{ for } D \to 1 ; ~ R \to \left( 1 - \delta \right) R^\prime \text{ for } D \to \infty . \label{scalingforinterpolation}
\end{equation}

An approximation for $D = 3$ (where $R = R^\prime$) emerges:

\begin{equation}
\epsilon_3 \left( R^\prime \right) = \frac{1}{3} \epsilon_1 \left( \frac{1}{3} R^\prime \right) + \frac{2}{3} \epsilon_\infty \left( \frac{2}{3} R^\prime \right), \label{scaledinterpolationformula}
\end{equation}
interpolating linearly between the dimensional limits  \cite{frantz1988, Tan_and_Loeser, lopez1993scaling, loeser1994correlated}.

For the random walk problem an important quantity is the principal components of the radius of gyration $R_i ^2$. These are one of the quantities for measuring the anisotropy of a $D$-dimensional random object. We propose to use the following  interpolation formula for random walk problem

\begin{equation}
(R_i^2)_{d} = \delta (R_i^2)_{1} + \left( 1- \delta \right) (R_i^2)_{\infty}, \label{interpolation_formula_for_rw}
\end{equation}
where $\delta = 1/D $, $D =$ dimension, and $(R_i^2)_{D}$ is the $i$-th principle component of the radius of gyration.

\section{\label{sec:Large-D limit}Random walk at the large-D limit}

There are a number of ways to measure the anisotropy of a random object generated by a three dimensional random walk. The shape of an arbitrary solid object in $D$-dimension is described by a quantity called the moment of inertia tensor, which is described as

\begin{equation}
T_{ij} = \frac{1}{N+1} \sum_{l=1}^{N+1} \left( x_{li} - \langle x_i \rangle \right) \left( x_{lj} - \langle x_j \rangle \right), \label{moment_of_inertia_tensor}
\end{equation} 
with
\begin{equation}
\langle x_i \rangle = \frac{1}{N+1} \sum_{l=1}^{N+1} x_{li},
\end{equation}
where $x_{li}$ is the $i$-th coordinate of the particle after $l$ steps and $N$ is the total number of steps.

Eigenvalues of $\overset\leftrightarrow{T}$ are the square of the components of the radius of gyration $R_i^2$. The gyration tensor was first introduced by Solc and Stockmeyer in their study of random flight chain \cite{vsolc1971shape}. These principal components determine the size and shape of a solid object and also inertial properties \cite{vsolc1973statistical}. By convention $R_1^2 \geq R_2^2 \geq R_3^2 ...$ . The combination of principal radii of gyration is known as the square of the radius of gyration 
\begin{equation}
R^2 = R_1^2 + R_2^2 + R_3^2 + ...
\end{equation}
The average value $\langle R ^2 \rangle$  for long, unrestricted open chain walks is calculated long ago, $R^2 = N/6$ \cite{debye1946intrinsic} and for closed walks $R^2 = N/12$ \cite{kramers1946behavior, Zimm1949}. From the principal radii of gyration we can compute asphericity or the asymmetry parameter of a solid object which is a one parameter measure to describe how much it is deformed from the perfect sphere \cite{theodorou1985shape, rudnick1986aspherity, aronovitz1986universal}.

For a $N$-step random walk problem at $D=\infty$,  Rudnick et al \cite{rudnick1987shapes} showed that the moment of inertia tensor becomes: 
\begin{eqnarray}
(T_\infty)_{ij} &=& \frac{i}{(N+1)^2} \left( N+1 -j\right) , \text{ for } i< j , \text{ and } 1\leq i, j\leq (N+1), \\ \nonumber
&=& \frac{j}{(N+1)^2} \left( N+1 -i\right) , \text{ for } i> j, \\ \nonumber
&=& 0 , \text{ otherwise. }
\end{eqnarray}
The eigenfunctions of $T_{ij}$ can be written as \cite{rudnick1987shapes2}:
\begin{equation}
\psi_n (i) = \left( \frac{2}{N+1}\right)^{1/2} \sin \left( \frac{n \pi i}{N+1}\right),
\end{equation}
and the corresponding eigenvalues are:
\begin{equation}
\lambda_n = (R_n)^2 =  \frac{1}{4(N+1)} \left[\sin^2 \left( \frac{n \pi }{2(N+1)}\right)\right]^{-1}, \label{eigen_values}
\end{equation} 
where $N$ is the total number of steps. For large $N$, the principal components of the radius of gyration becomes
\begin{equation}
\langle R_n ^2 \rangle = \lambda_n  = \frac{N+1}{\pi^2 n^2} \approx \frac{N}{\pi^2 n^2}.
\end{equation}

\section{\label{sec:one dim} Random walk in one dimension }

In one dimension, we have only two possible directions: forward and backward. For total number of steps = $N$, the probability of $m$ forward steps and $n$ backward steps can be written as
\begin{equation}
P(m,n) = \frac{1}{2^N}\frac{N !}{m! n!} .
\end{equation}
For large $N$ the probability function can be defined as:
\begin{equation}
P(x) = \frac{2}{\sqrt{2 \pi N}} e^{-\frac{x^2}{2N}},
\end{equation}
where $x$ is the distance from the origin.
\begin{figure}[H]
  \centering
 \includegraphics[width=0.6\textwidth]{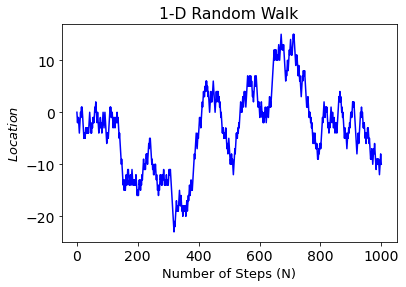}
  \caption{A random walk in one dimension with $1000$ steps. The x-axis describes the number of steps and the y-axis describes the position of the random walker after each step from the origin. The forward and backward directions are assumed to be positive and negative respectively.}
\label{figure1}
\end{figure}

We have developed a python program to simulate a random walk in one-dimension. For $N = 1000$ steps an example of one dimensional random walk is plotted above in Fig.\ref{figure1}.

We calculate the moment of inertia tensor $T$ from Eq.(\ref{moment_of_inertia_tensor}) which has one eigenvalue, $R_1^2$, the principle component of the radius of gyration. We repeat the simulation for $100,000$ times and take the average of all $R^2$'s and obtain $\langle R_1 ^2 \rangle = 166.2851794$.

The distribution of all $R_1^2$s obtained from different simulations is plotted below
\begin{figure}[H]
  \centering
 \includegraphics[width=0.8\textwidth]{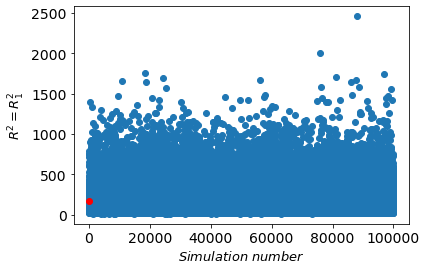}
  \caption{The distributio of all $R^2$s obtained from different simulations. The red point is the average $\langle R_1 ^2 \rangle$ .}
\label{figure2}
\end{figure}

\section{\label{sec:Two dim} Random walk in two dimension}
In two dimensions the particle can go in any of the two directions X and Y. We simulate a random walk in two spacial dimensions with $N = 1000$ steps starting from the origin $(0,0)$. The simulation result of a two dimensional random walk is plotted below 

\begin{figure}[H]
  \centering
 \includegraphics[width=0.7\textwidth]{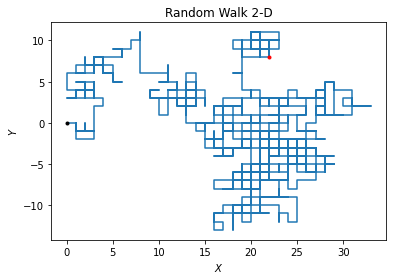}
  \caption{A random walk in two dimension with 1000 steps. The blue and red dots are beginning (origin) and ending points respectively.}
\label{figure32}
\end{figure}

We calculate the moment of inertia tensor $T_{ij}$ from Eq.(\ref{moment_of_inertia_tensor}), which has two eigenvalues, $R_1^2, R_2^2$, the principle components of the radius of gyration. By convention $R_1^2 \geq R_2^2$. We repeat the simulation for $100,000$ times. The distribution of all $R_1^2, R_2^2$s obtained from different simulations is plotted below
\begin{figure}[H]
  \centering
 \includegraphics[width=0.7\textwidth]{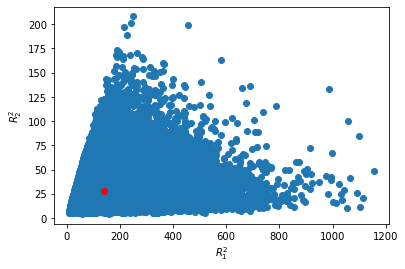}
  \caption{The distribution of all $R_1^2, R_2^2$s obtained from different simulations. The red point is the average $\langle R^2 \rangle$.}
\label{figure42}
\end{figure}
The averages of all $R_i^2$'s are given by $\langle R_1 ^2 \rangle = 138.94300$,  $\langle R_2 ^2 \rangle = 27.90438$. The radius of gyration is given by: $R^2 = (R_1^2 + R_2^2) = 166.84739$.

\begin{figure}[H]
    \centering
    \begin{tikzpicture}[
 image/.style = {text width=0.5\textwidth, 
                 inner sep=0pt, outer sep=0pt},
node distance = 1mm and 1mm
                        ] 
\node [image] (frame1)
    {\includegraphics[width=\linewidth]{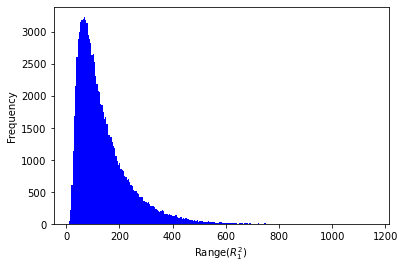}};
\node [image,right=of frame1] (frame2) 
    {\includegraphics[width=\linewidth]{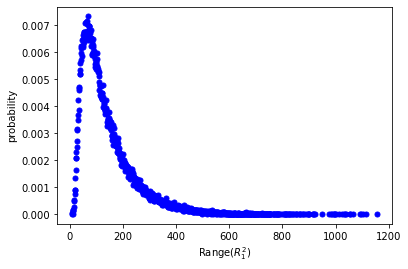}};
\end{tikzpicture}
    \caption{{plots of frequency and probability of $ R_1 ^2$, as a function of its values occurring during different simulation.}}
    \label{figure62}
\end{figure}

\section{\label{sec:Three dim}Random walk in three dimension}
In three dimensions the particle can go in any of the three X, Y, and Z directions. 

\begin{figure}[H]
  \centering
 \includegraphics[width=0.6\textwidth]{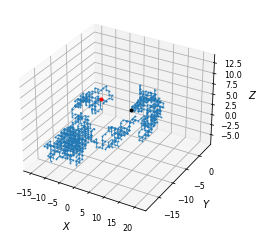}
  \caption{A random walk in three dimension with 1000 steps. The blue and red dots are beginning (origin) and ending points respectively.}
\label{figure3}
\end{figure}

\begin{figure}[H]
  \centering
 \includegraphics[width=0.5\textwidth]{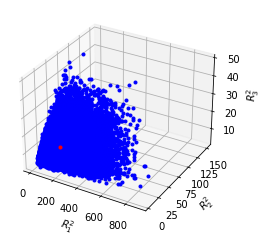}
  \caption{The distribution of all $(R_1^2, R_2^2, R_3^2)$s obtained from different simulations. The red point is the average $\langle R^2 \rangle$.}
\label{figure4}
\end{figure}

\begin{figure}[H]
    \centering
    \begin{tikzpicture}[
 image/.style = {text width=0.5\textwidth, 
                 inner sep=0pt, outer sep=0pt},
node distance = 1mm and 1mm
                        ] 
\node [image] (frame1)
    {\includegraphics[width=\linewidth]{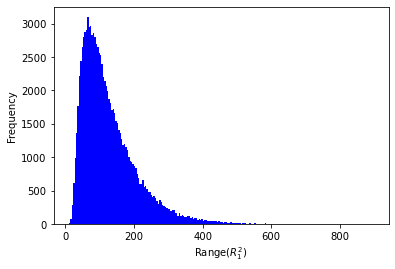}};
\node [image,right=of frame1] (frame2) 
    {\includegraphics[width=\linewidth]{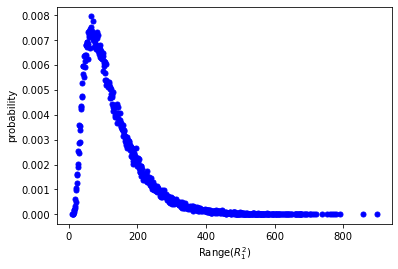}};

\end{tikzpicture}
    \caption{Frequency and probability distribution of $ R_1 ^2$.}
    \label{figure6}
\end{figure}

We simulate a random walk in three spacial dimensions with $N = 1000$ steps starting from the origin $(0,0,0)$. The simulation result of a three dimensional random walk is plotted above in Fig.\ref{figure3}. We calculate the moment of inertia tensor $T_{ij}$ from Eq.(\ref{moment_of_inertia_tensor}), which has three eigenvalues, $R_1^2, R_2^2, R_3^2$, the principle components of the radius of gyration. By convention $R_1^2 \geq R_2^2 \geq R_3^2$. We repeat the simulation for $100,000$ times. The distribution of all $(R_1^2, R_2^2, R_3^2)$s obtained from different simulations is plotted in Fig.\ref{figure4} and the frequency and probability distribution of the $R_1^2$ is plotted in Fig.\ref{figure6}. 

The averages of all $R_i^2$'s are given by $\langle R_1 ^2 \rangle = 126.82723$,  $\langle R_2 ^2 \rangle = 28.68618$, $\langle R_3 ^2 \rangle = 10.57850$. The radius of gyration is given by: $R^2 = (R_1^2 + R_2^2 + R_3^2) = 166.0919$.


\section{\label{sec:Interpolation3D} Interpolation at $D=3$}

We employ the interpolation formula in Eq.(\ref{interpolation_formula_for_rw}) developed by Herschbach \cite{Herschbach2017} for $R_i^2$ in three-dimension as follows:

\begin{equation}
(R_i^2)_{3} = \frac{1}{3} (R_i^2)_{1} + \frac{2}{3} (R_i^2)_{\infty}, \label{dim_int_three_dim}
\end{equation}
where $(R_i^2)_{D}$ is the $i$-th component of the radius of gyration in D-dimension.

From Eq.(\ref{eigen_values}) we see that

\begin{equation}
(R_n)^2 =  \frac{1}{4(N+1)} \left[\sin^2 \left( \frac{n \pi }{2(N+1)}\right)\right]^{-1}, \text{ for } m=1,2,3, \text{ and } N= 1000. \label{inf_dim_r_i}
\end{equation} 



Therefore from Eq.(\ref{inf_dim_r_i}) $(R_1^2)_\infty = 101.422588$. On the other hand $(R_1^2)_1 = 166.2851794$.


From the above Eq. (\ref{dim_int_three_dim}) we can calculate
\begin{equation}
(R_1^2)_{3} = \frac{1}{3} (R_1^2)_{1} + \frac{2}{3} (R_1^2)_{\infty} = {123.04345}.
\end{equation}
The computer simulation of the three-dimensional random walk gives ${(R_1^2)_{sim} = 126.82723}$, which is a very close estimation with $2 \%$ error. 

In one dimension, there is only one component of the radius of gyration. Therefore, for interpolating the other component of the radius of gyration we have extrapolate the values of $R_n^2$ at one dimension for the calculation purpose. For $D = \infty $, and large $N$, the principal components of the radii are gyration are as follows:
\begin{equation}
R_n^2 = \frac{1}{n^2} R_1^2, 
\end{equation}
so that 
\begin{equation}
(R_1^2)_{\inf}:  (R_2^2)_{\inf}: (R_3^2)_{\inf} = 9 : 4 : 1. \label{3dratiosinf}
\end{equation}

On the other hand at $D=3$  in article \cite{vsolc1973statistical} the authors showed that the following limiting ratios have been found to be approached for the large $N$ limit
\begin{equation}
\langle R_1 ^2 \rangle : \langle R_2 ^2 \rangle : \langle R_3 ^2 \rangle = 11.80 : 2.69 : 1 \label{3dratios}
\end{equation}

From the above Eq.s (\ref{3dratiosinf}) and (\ref{3dratios}), we see that the ratios between the radius of gyration changes with the change of dimension. We propose a model for the dimensional dependence of $R_n ^2$ as follows:

\begin{equation}
R_n^2 = \left(\frac{1}{n^2(1 + \frac{a_n}{D})} \right) R_1^2, \text{ for } n= 2, 3,...,
\end{equation}
where $a_2, a_3, ...$ are constants and $D$ is the dimension. Now from the standard result at $D =3$ from Eq.(\ref{3dratios}) we can get two sets of equations, which has a solution $a_2= 0.28996$ and $a_3 = 0.9333$.

With this ratios we can calculate the ratios between $R_n^2$ at $D=1$ 

\begin{equation}
(R_2^2)_1 = \frac{1}{5.15985} (R_1^2)_1 , ~ \text{ and } ~ (R_3^2)_1 = \frac{1}{17.39997} (R_1^2)_1 ,  \label{eeeq1dim}
\end{equation}
%
%
%

For $N = 1000$, from Eq. (\ref{eeeq1dim}), at $D=1$, we can calculate the following quantities $(R_2^2)_{1} =32.2267 $ and $(R_3^2)_{1} = 9.5566$.
And from Eq. (\ref{inf_dim_r_i}) at $D=\infty$, we compute the following quantities $(R_2^2)_{\infty} = 25.3557$ and $(R_3^2)_{\infty} = 11.26925$. 
With the values of $R_i^2$ at $D=1$ and $D=\infty$, we  calculate $(R_2^2)_{3} = {27.64604}$ and the computer simulation gives $(R_2^2)_{sim} = {28.68618}$. Whereas, $(R_3^2)_{3} = {10.6983}$ and the computer simulation gives $(R_3^2)_{sim} = {10.5785}$.


The total radius of gyration from the interpolation is given by 
\begin{equation}
(R^2)_{3} = (R_1^2)_{3}+(R_2^2)_{3}+(R_3^2)_{3} = {161.38},
\end{equation}
which is a fairly good estimation for the $(R^2)_{sim} = {166.0919}$, obtained from random walk simulation at $D=3$. 

Now, we also calculate the ratios between the radii of gyrations obtained from the interpolation: 
$(R_1^2)_{3} : (R_2^2)_{3} : (R_3^2)_{3} = 11.5: 2.58 : 1 $, 
where as ratios between the radii of gyrations obtained from simulation is given by
$(R_1^2)_{sim} : (R_2^2)_{sim} : (R_3^2)_{sim} = 11.98: 2.71 : 1 $. 


The above estimates from dimensional interpolation are fairly close estimates, because none of the random number generators are perfect random number generator. Therefore the computer simulation of the random walk at $D=3$ is only an approximation for a perfect random walk in three dimensions.
We expand our calculation by considering different step sizes to establish the validity of our interpolation formula. The averages are taken over $1000$ samples.

\textbf{ N= 10}\\
Three dimensional simulation of all $R_i^2$'s are given by $\langle R_1 ^2 \rangle = 1.3640$,  $\langle R_2 ^2 \rangle = 0.3242$, $\langle R_3 ^2 \rangle = 0.1139$.
On the other hand from $D=\infty$ we have $(R_1^2)_\infty = 1.1221$ and from $D=1$ we have $(R_1^2)_1 = 1.8226$.
Using Eqs (\ref{inf_dim_r_i}, \ref{eeeq1dim}) with the  interpolation formula in Eq. (\ref{dim_int_three_dim}) we can calculate $(R_1^2)_{3} = {1.35563}$, which is close to the value $\langle R_1 ^2 \rangle = {1.3640}$.
Like the previous case we calculate the value of $R_2^2$ from dimensional interpolation which is equal to $(R_2^2)_{3} = 0.3086$, which agrees well with the computer simulation $(R_2 ^2)_{sim} = 0.3242$. For the third component $(R_3^2)_{3} = 0.1142$, which agrees well with the computer simulation $(R_3 ^2)_{sim} = 0.1140$. 

\textbf{ N= 50}\\
Three dimensional simulation of all $R_i^2$'s are given by $\langle R_1 ^2 \rangle = 6.3367$,  $\langle R_2 ^2 \rangle = 1.4336$, $\langle R_3 ^2 \rangle = 0.54268$.
On the other hand from $D=\infty$ we have $(R_1^2)_\infty = 5.1690$ and from $D=1$ we have $(R_1^2)_1 = 8.29468$.
Using Eqs (\ref{inf_dim_r_i}, \ref{eeeq1dim}) with the  interpolation formula in Eq. (\ref{dim_int_three_dim}) we can calculate $(R_1^2)_{3} = {6.2109}$, which is close to the value $\langle R_1 ^2 \rangle = {6.3367}$.
Like the previous case we calculate the value of $R_2^2$ from dimensional interpolation which is equal to $(R_2^2)_{3} = 1.3981676$, which agrees well with the computer simulation $(R_2 ^2)_{sim} = 1.4336$. For the third component $(R_3^2)_{3} = 0.5428$, which agrees well with the computer simulation $(R_3 ^2)_{sim} = 0.5427$.



\textbf{ N= 100}\\
Three dimensional simulation of all $R_i^2$'s are given by $\langle R_1 ^2 \rangle = 12.7003$,  $\langle R_2 ^2 \rangle = 2.9590$, $\langle R_3 ^2 \rangle = 1.0776$. 
On the other hand from $D=\infty$ we have $(R_1^2)_\infty = 10.2342$ and from $D=1$ we have $(R_1^2)_1 = 17.5066$.
Using Eqs (\ref{inf_dim_r_i}, \ref{eeeq1dim}) with the  interpolation formula in Eq. (\ref{dim_int_three_dim}) we can calculate $(R_1^2)_{3} = {12.6584}$, which is close to the value $\langle R_1 ^2 \rangle = {12.7003}$.
Like the previous case we calculate the value of $R_2^2$ from dimensional interpolation which is equal to $(R_2^2)_{3} = 2.83707$, which agrees well with the computer simulation $(R_2 ^2)_{sim} = 2.9590$. For the third component $(R_3^2)_{3} = 1.0939$, which agrees well with the computer simulation $(R_3 ^2)_{sim} = 1.0776$.

\section{\label{sec:Interpolation2D}Interpolation at $D=2$}

We employ the interpolation formula from Eq.(\ref{interpolation_formula_for_rw}) for $R_i^2$ in three-dimension as follows:

\begin{equation}
(R_i^2)_{2} = \frac{1}{2} (R_i^2)_{1} + \frac{1}{2} (R_i^2)_{\infty}, \label{dim_int_two_dim}
\end{equation}
where $(R_i^2)_{D}$ is the $i$-th component of the radius of gyration in D-dimension.

%



Therefore from Eq.(\ref{eigen_values}), $(R_1^2)_\infty = 101.422588$. On the other hand $(R_1^2)_1 = 166.28518$.


From the above Eq. (\ref{dim_int_two_dim}) we can calculate
\begin{equation}
(R_1^2)_{2} = \frac{1}{2} (R_1^2)_{1} + \frac{1}{2} (R_1^2)_{\infty} = {133.8539}.
\end{equation}
From the computer simulation of the two-dimensional random walk gives ${(R_1^2)_{sim} = 138.94300}$, which is a fair estimation. 

Using Eq. (\ref{eeeq1dim}) we write the ratio between $R_2^2$ and $R_1^2$ at $D=1$ 

\begin{equation}
(R_2^2)_1 = \frac{1}{5.15985} (R_1^2)_1 .  \label{eeeq1dim2}
\end{equation}
%
%
%

For $N = 1000$ from Eq. (\ref{eeeq1dim2}), at $D=1$, we calculate $(R_2^2)_{1} =32.2267 $.
From Eq. (\ref{inf_dim_r_i}) at $D=\infty$, we calculate the following quantities $(R_2^2)_{\infty} = 25.3557$.
With the above data at $D=1$ and $D=\infty$, we compute $(R_2^2)_{2} = {28.7912}$ and the computer simulation gives $(R_2^2)_{sim} = {27.9044}$. 

The total radius of gyration from the interpolation is given by 
\begin{equation}
(R^2)_{2} = (R_1^2)_{2}+(R_2^2)_{2} = {162.6451},
\end{equation}
which is a fairly good estimation for the $(R^2)_{sim} = {166.84739}$, obtained from random walk simulation in two dimensions. 

Like the three dimensional case, we also calculate the ratios between the radii of gyrations obtained from the interpolation: 
$(R_1^2)_{2} : (R_2^2)_{3}  = 133.8539:28.7912=4.65: 1 $, 
whereas ratios between the radii of gyrations obtained from simulation is given by
$(R_1^2)_{sim} : (R_2^2)_{sim}  = 138.9430: 27.90438= 4.97: 1 $. 
%
We expand our calculation by considering different step sizes to establish the validity of our interpolation formula at $D=2$. The averages are taken over $1000$ samples.

\textbf{ N= 50}\\
Two dimensional simulation of all $R_i^2$'s are given by $\langle R_1 ^2 \rangle = 7.2121$,  $\langle R_2 ^2 \rangle = 1.4513$.
On the other hand from $D=\infty$ we have $(R_1^2)_\infty = 5.1690$ and from $D=1$ we have $(R_1^2)_1 = 8.29468$.
Using Eqs (\ref{inf_dim_r_i}, \ref{eeeq1dim}) with the  interpolation formula in Eq. (\ref{dim_int_three_dim}) we can calculate $(R_1^2)_{3} = {6.7318}$, which is close to the value $\langle R_1 ^2 \rangle = {7.2121}$.
Like the previous case we calculate the value of $R_2^2$ from dimensional interpolation which is equal to $(R_2^2)_{2} = 1.4505$, which agrees well with the computer simulation $(R_2 ^2)_{sim} = 1.4513$. 



\textbf{ N= 100}\\
Two dimensional simulation of all $R_i^2$'s are given by $\langle R_1 ^2 \rangle = 14.0010$,  $\langle R_2 ^2 \rangle = 2.8198$.
On the other hand from $D=\infty$ we have $(R_1^2)_\infty = 10.2342$ and from $D=1$ we have $(R_1^2)_1 = 17.5066$.
Using Eqs (\ref{inf_dim_r_i}, \ref{eeeq1dim}) with the  interpolation formula in Eq. (\ref{dim_int_three_dim}) we can calculate $(R_1^2)_{2} = {13.8704}$, which is close to the value $\langle R_1 ^2 \rangle = {14.0010}$.
Like the previous case we calculate the value of $R_2^2$ from dimensional interpolation which is equal to $(R_2^2)_{2} = 2.9760$, which agrees well with the computer simulation $(R_2 ^2)_{sim} =  2.8198$. 

\section{\label{sec:asphericity}Asphericity}

The radius of gyration is a measure of the average extent of a random walk. However, to get a better idea of its shape a quantity $A_D$ is defined, named the asphericity \cite{theodorou1985shape,rudnick1987shapes2, aronovitz1986universal}. This quantity determines how much a solid object is deviated from the perfect sphere. For example for a perfect spherical object asphericity $A_D = 0$, on the other hand this quantity has an upper bound of one, a limit that is reached when the walk is extended in one dimension only. Mathematically expression for $A_D$ is as follows:
\begin{equation}
A_D = \frac{\sum_{i>j}^D \left\langle \left( R_i^2 - R_j^2 \right)^2 \right\rangle}{\left( D-1 \right) \left\langle \left( \sum_{i=1}^D R_i^2 \right)^2 \right\rangle} \label{asphericity}
\end{equation}

At the large $D$-dimension Rudnick et. al. \cite{theodorou1985shape,rudnick1987shapes2}  showed that the expression of  asphericity can be written as 

\begin{equation}
A_D = \frac{2}{5} + \frac{12}{25 D} + O\left(\frac{1}{D^2}\right). \label{series}
\end{equation}

For $D \to \infty$, $A_\infty = \frac{2}{5}$. On the other hand, for $D=1$, $A_1 = 1$. 

In \cite{tsipis1996new, herschbach1996dimensional} the authors introduces a dimensional interpolation formula  with $D=0$ and $D=\infty$ and calculate the  a quantity called asymmetry (similar to asphericity), which is defined as

\begin{equation}
\mathcal{A}_D = \frac{1}{ D-1 } \left\langle\frac{\sum_{i>j}^D  \left( R_i^2 - R_j^2 \right)^2}{\left( \sum_{i=1}^D R_i^2 \right)^2} \right\rangle. \label{asymmetry}
\end{equation}

Although both the above quantities measure the anisotropy, the asymmetry $\mathcal{A}_D $ is slightly different from the asphericity $A_D$, defined in Eq. (\ref{asphericity}). Because,  the asphericity of each walk in the ensemble calculated first and then the result is  averaged. Whereas, for calculating the asphericity $A_D$ we take the average over the numerator and denominator part separately.  See \cite{rudnick1987shapes, rudnick1986aspherity} for more details.

Eq. (\ref{series}) gives a dimensional dependence of $A_D$ at the large-$D$ limit as a power series of $1/D$. Although this power series expression for $A_D$ does not produce the result at $D=1$, which is $A_1 =1$. For the interpolation at $D=3$ we modify the above expression in Eq. (\ref{series}) for $A_D$ for the large-$D$ limit,  and use the result for $A_D$ at $D=1$ . We rewrite the expression  for the asphericity in $D$-dimension:

\begin{equation}
A_D = \frac{2}{5} + \frac{a}{D} +\frac{a^2}{D^2} + \frac{a^3}{D^3}+ ..., \label{newseries}
\end{equation}

where $a$ is a parameter to be fixed from a known value of $A_D$ at a given dimension. For the calculation purpose we consider the terms up to the cubic order in $(a/D)$. For $D=1$, we know that $A_D =1$. We put this $D=1$ result in the above Eq. (\ref{newseries}) and get the following condition for $a$

\begin{equation}
5 a^3 +5a^2+5a=3,
\end{equation}

which has a solution $a=0.3894$. Now, putting back the value of $a$ and $D=3$ in Eq.(\ref{newseries}) we obtain 
\begin{equation}
A_3 = 0.5488.
\end{equation}

We have performed numerical simulation of  a random walk at  $D = 3$ with $N = 1000$ steps. Then, we took an average over $100000$ samples and calculate the asphericity using Eq. (\ref{asphericity}) $A_{sim}= 0.5240$, which is in close estimation with the above theoretical prediction.

\section{\label{sec:Conclusion}Conclusion}

The simplicity of the $D \to \infty$ limit attributed to the removal of the derivatives terms from the Hamiltonian in the case of electronic structure \cite{loeser1994correlated}. Thus one has to find the minimum energy of an effective potential.  For the random walk, at $D \to \infty$ limit, the walker takes a step in along each $D$-dimensional axis.  
The simplicity of the $D = 1$ limit keeps derivatives in a Hamiltonian and is a true hyperquantum limit.  Combining these extreme partner limits delivers the dimensional interpolation formula.  
We have used the interpolation formula successfully for two-electron atoms \cite{Herschbach2017} and generalized it for few electron atoms, simple diatomic molecules \cite{ghosh2020unorthodox}, and metallic hydrogen\cite{metallic_hydrogen}.   In this article we implemented the dimensional interpolation formula for the random walks and calculated physical quantities like radii of gyration and asphericity in two and three dimensions to show its robustness in different topics in physics and chemistry.

The complexity of $N$-step random walk problem at $D=3$ is of order order $6^N$, where as in $D=1$, it is of the order $2^N$. Therefore, in $D=3$ the complexity of the simulation grows exponentially $(6^N/2^N = 3^N)$ with the number of steps and with the number of dimension. Whereas, for random walks at $D=\infty$ there is an analytical formula, which is described above.  It is relatively easy to calculate the $D \to \infty$ and $D = 1$ limits, so the interpolation formula can predict results for the physical dimensions, $D = 2 \text{ or } 3$. The interpolation formula is general and might be used to obtain accurate results for other complex systems such as spin systems on different latices used extensively for magnetic materials and quantum computing simulations.

\section*{Authors' information}
ORCID IDs: Kumar Ghosh: 0000-0002-4628-6951, Sabre Kais: 0000-0003-0574-5346, Dudley Herschbach: 0000-0003-3225-0648. Note: The authors have no competing financial interest.

\section*{Acknowledgements}
S. K. acknowledges funding by the U.S. Department of Energy (Office of Basic Energy Sciences) under Award No. DE-SC0019215.

\bibliography{ref}

\end{document}